\documentclass[11pt,onecolumn,amssymb,nofootinbib]{revtex4}
\usepackage{amsmath, amsthm, amscd, amssymb}
\usepackage{graphicx, braket}
\usepackage{bm}
\usepackage{bbm}
\usepackage{epstopdf}

\begin{document}

\title{\bf A One-Dimensional Model for Star Formation Near a ``Leaky'' Black Hole}

\author{Stephen L. Adler}
\email{adler@ias.edu} \affiliation{Institute for Advanced Study,
Einstein Drive, Princeton, NJ 08540, USA.}

\author{Kyle Singh}
\email{kylesi@sas.upenn.edu}\affiliation{Department of Physics, University of Pennsylvania, Philadelphia, PA 1910}
\begin{abstract}
In the presence of a Weyl scaling invariant cosmological action, black holes no longer have an event horizon and an apparent horizon. So they have no trapped surfaces, and may be ``leaky'', emitting a ``black hole wind'' which could lead to star formation in the neighborhood of the hole.  In this paper we formulate and analyze a one-dimensional model for star formation resulting from a postulated outgoing black hole particle flux, incident on a  distant spherical surface modeled as a set of planar disks surrounding the hole.  Using the Toomre analysis of the Jeans
instability of a disk, we compute conditions for a disk collapse instability and estimate the collapse time. We suggest a mechanism for giving the disk angular momentum around the central black hole.   This gives a possible explanation for  the puzzling observation of young stars forming in the vicinity of the black hole Sagittarius A* central to the Milky Way galaxy.
\end{abstract}

\maketitle
\section{Introduction}

A series of papers by one of the authors (SLA), recently reviewed in \cite{adler1}, have studied the implications for cosmology and for black hole properties of a Weyl scaling invariant
dark energy action
\begin{equation}\label{dark}
S_{\rm eff}=-\frac{\Lambda}{8 \pi G} \int d^4x ({}^{(4)}g)^{1/2}(g_{00})^{-2}~~~,
\end{equation}
where $g_{\mu\nu}$ is the spacetime metric, $\Lambda$ is the observed cosmological constant, and $G$ is the Newton gravitational constant.
Specifically, with this dark energy action, black holes are modified by having no event horizon and no
apparent horizon, so that in principle material can leak out from the hole, giving rise to a black hole ``wind'' \cite{adler2}.  As a possible astrophysical application, we noted in \cite{adler2} that
  observations of Sgr A*, the central  black hole in our galaxy,  show the presence of young stars in its close vicinity.  Lu et al. open their article \cite{lu} by stating that: ``One of the most perplexing problems associated with the supermassive black hole at the center of our Galaxy is the origin of young stars in its close vicinity''.  Similar clusters of young stars are found in the vicinity of  supermassive black holes in nearby galaxies \cite{lu1}.   A black hole wind of particles could possibly furnish a mechanism for
    the origin of such stars, both by providing material for their formation, and by providing an outward pressure cushioning nascent stars  against the gravitational tidal forces of the black hole.  We suggested in \cite{adler2} that this idea could be tested in a phenomenological way, by postulating parameters for the black hole wind (particle type, velocity, flux) and seeing if they can account for young star formation.  The purpose of the present article is to follow up on this suggestion with a simple one dimensional model.

\section{One Dimensional Model}
\subsection{Formulation of the Model}
The young stars near Sgr A* are situated at a radius $R* \simeq 0.42 \times 10^{13}$ km from the hole, which has a nominal horizon radius of around $10^7$ km.  So the spherical shell containing the young stars has, from the vantage point of the hole, a very large radius of curvature. To make a simplified model for star formation, we start by ignoring three dimensional aspects of the geometry, by approximating a segment of the spherical shell as a planar disk, with the shell the union of such segments.   A radial flux of particles emanating from the hole is then replaced by a flux of particles along the normal to the disk, which we label as the $z$ axis.

This simplification corresponds to the following model.  Consider a flux $f$ of particles of mass $m$ moving upwards along the $z$ axis, with initial velocity
distribution $w(v_0)$ at radius $z_0$, in the presence of a gravitational potential $V(z)=mgz$. For the moment, we neglect a possible infalling flux of particles arising from $z=\infty$, but later on we will see that we have to include it.
For the gravitational acceleration $g$ to mock up
the radial acceleration near the young star formation region, we take
\begin{equation}\label{gdef}
g=GM/(R*)^2~~~,
\end{equation}
with  $M$ the black hole mass, which for Sgr A* is around $4.3 \times 10^6 M_\odot$ in terms of the solar mass $M_\odot$.  The  velocity distribution function $w(v_0)$ is not known, but to illustrate its effect we assume that it has a Maxwellian form,
\begin{equation}\label{vdist}
w(v_0)= \frac{4 B^{3/2}}{\pi^{1/2}}v_0^2 e^{-B v_0^2}~~~,
\end{equation}
normalized so that
\begin{equation}\label{norm}
\int_0^\infty dv_0 w(v_0) =1~~~.
\end{equation}
\subsection{Noninteracting particles}
 To start analyzing the model, let us first assume that the flux particles are not interacting, either gravitationally or through collisions.
 Then since the total energy, kinetic plus potential, is conserved, the velocity $v(z)$ at height $z$ is related to the velocity $v_0$ at height $z_0$ by the formula
\begin{equation}\label{encons}
E=\frac{1}{2}mv_0^2 +mgz_0 = \frac{1}{2} mv^2 +mgz
\end{equation}
giving
\begin{equation}\label{vatz}
v(z)=[v_0^2-2g(z-z_0)]^{1/2} ~~~.
\end{equation}
At a maximum height $z_{\rm max}(v_0)=z_0 + v_0^2/(2g)$ the particles have zero upwards velocity and reverse motion to fall back down to $z=z_0$. In terms of the
particle flux $f$ and velocity $v$, the particle density $n(z)$ at a given $z$ is given by
\begin{align}\label{dens}
n(z)=&f/v(z)\cr
=& f/[v_0^2-2g(z-z_0)]^{1/2} ~~~\rm{when}~~z\leq z_{\rm max}~~~,\cr
=& 0 ~~~\rm{when}~~ z>z_{\rm max} ~~~.
\end{align}
Thus if the velocity distribution $w(v_0)$ were a delta function $w(v_0)=\delta(v_0-v^*)$,
the particle density would have a cusp-like infinity at $z_{\rm max}(v^*)$. Taking $z_0=0$ as the origin of the $z$ axis, for a general velocity distribution $w(v_0)$ this cusp-like density
distribution is smeared into
\begin{equation}\label{smeared}
n(z)_{\rm AV}=f\int_{(2gz)^{1/2}}^\infty   dv_0 \frac{w(v_0)}{[v_0^2-2gz]^{1/2}}~~~,
\end{equation}
which for the Maxwellian velocity distribution of Eq. \eqref{vdist} becomes (writing $A^2=2gz$, and making a change of variable $v_0^2-A^2=x$)
\begin{equation}\label{smeared1}
n(z)_{\rm AV}=f\frac{2 B^{3/2}}{\pi^{1/2}} e^{-B A^2} \int_0^\infty   dx \frac{(x+A^2)^{1/2}}{x^{1/2}} e^{-Bx}~~~.
\end{equation}
This integral can be evaluated in terms of the complementary hypergeometric function ${\rm HypergeometricU}(a,b,c)$ to give
\begin{equation}\label{smeared2}
n(z)_{\rm AV}=f2B^{1/2} e^{-B A^2}{\rm HypergeometricU}(-1/2,0,A^2 B) ~~~.
\end{equation}
In Fig. 1 we have plotted the cusp distribution of Eq. \eqref{dens} for $v^*=B^{-1/2}$, which is the maximum of the Maxwellian distribution of Eq. \eqref{vdist}, together with the corresponding smeared distribution of Eq. \eqref{smeared2}, for illustrative values $f=1$ and $B=1.5$

\subsection{Including flux particle interactions}

Assuming that the flux particles are electrically neutral, there will be two types of interactions between them, short range scattering from one another (which can be modeled as hard sphere scattering), and long range gravitational interactions.\footnote{If the particles are electrically  charged, there will also be long range electromagnetic interactions and plasma complexities, which we ignore.}   Flux particle interactions will have two effects.  First, short range scattering will increase the density in the overdensity regions of Fig. 1, since some incoming flux particles will scatter into the overdensity region rather than falling back down, with the amount of such scattering increasing with the density of particles already present.    One expects the density integrated over the z axis to increase as a sigmoid or logistic function, first growing with an exponential contribution in time, then later growing  with a  rate leveling off to linear when the overdensity region becomes opaque to incident flux particles.  The gravitational interaction will result in the overdensity band shrinking in
width with time as a result of the attractive long range interaction.  So we expect to end up with an overdensity band that in first approximation can be considered a planar surface of small width along the $z$ axis compared to its extent normal to the $z$ axis, characterized by the total particle number, integrated along the z direction, per unit area of the planar surface.  Analytic modeling of the initial stage of the development of the overdensity surface will be complex, but it should be possible to study it by a mulltiparticle  simulation.

\subsection{Quasi-equilibrium of a planar layer}

We now assume that the overdensity region has evolved to take the form of a thin planar layer, with particle number surface density $\Sigma$, corresponding to a mass surface density $m\Sigma$, and equilibrium temperature $T$. The upward moving ``wind'' particles incident on the lower surface of this layer are characterized by  the flux $f$ and root mean square (rms) velocity $v_{\rm w}$.   We make the following preliminary assumptions about the overdensity region:    (i) It is optically thick to the flux of incoming
particles, so that they are all absorbed. (ii)  It is supported from falling down in  the gravitational potential (i.e., from falling  into the black hole in the original three dimensional geometry) by the pressure arising from the momentum gain rate from the absorbed particle flux. (iii)  It is approximately optically thick to electromagnetic radiation, so that we can use the Stefan-Boltzmann law to calculate energy emission from the planar surfaces, with emissivity $\epsilon \sim 1$.\footnote{If included in the subsequent formulas, $\epsilon$ would appear raised to the $-1/4$ or $-1/8$ power, so our results are relatively insensitive to the value of the emissivity.}   (iv) It's temperature is fixed by  balancing this energy loss rate with the energy gain rate from the kinetic energy of the absorbed particle flux.

According to assumption (i), and neglecting leakage of particles out of the lower and upper surfaces of the overdensity layer, the particle number surface density $\Sigma$ increases with time from absorption of the incident flux as
\begin{equation}\label{incr}
\Sigma(t)=\Sigma(0)+ ft~~~.
\end{equation}
The effective weight of a unit area of surface is $mg \Sigma(t) $, and by assumption (ii) this is supported by the absorbed momentum per unit area per unit time of the particles incident on the lower surface, which is $m v_{\rm AV} f$.  Neglecting the difference between the average
velocity $v_{\rm AV}$ and the rms velocity $v_{\rm w}$, the condition for the surface to be supported is
\begin{equation}\label{supp}
g \Sigma(t)=v_{\rm AV}f \simeq v_{\rm w} f~~~.
\end{equation}
Recalling Eq. \eqref{vatz},  this implies that as $\Sigma(t)$ increases through the absorption of particles,  the overdensity layer sinks downwards to a region with larger $v_{\rm w}$.

The rate of energy increase of the overdensity layer,  from absorption of the kinetic energy of the incident particles, is $(1/2)mv_{\rm w}^2 f$.
According to assumptions (iii) and (iv), this is to be balanced against the rate of blackbody radiation from the two surfaces of the overdensity region, which assuming emissivity of order unity  is $2\times \sigma T^4$, with $\sigma= (\pi^2/60) k^4/(c^2 \hbar^3)$ the
Stefan-Boltzmann constant, expressed in terms of the Boltzmann constant  $k$  ,  the velocity of light   $c$, and  the reduced Planck constant  $\hbar$.   Assuming an ideal gas law $m v_{\rm rms}^2 =3kT$ relating the temperature $T$ to the rms velocity of the particles $v_{\rm rms}$ in the overdensity region (not to be confused with the incident wind rms velocity $v_{\rm w}$!), the energy balance equation
\begin{equation}\label{enbal1}
\frac{1}{2}m v_{\rm w}^2 f = 2 \sigma T^4 = \frac{\pi^2}{30}\frac{(kT)^4}{c^2\hbar^3}~~~
\end{equation}
gives  formulas for $v^2_{\rm rms}$ and $T$,
\begin{align}\label{enbal2}
v^2_{\rm rms}=&3 m^{-3/4} [(15/\pi^2)v_{\rm w}^2 f c^2 \hbar^3]^{1/4}~~~,\cr
T=&k^{-1}[(15/\pi^2)mv_{\rm w}^2 f c^2 \hbar^3]^{1/4}~~~.\cr
\end{align}
This completes the quasi-equilibrium  conditions governing the state of the overdensity region, allowing us next to determine the stability
of this region against gravitational collapse.
\subsection{Jeans criterion for a thin disk}

To assess the stability of the overdensity layer against gravitational collapse, we visualize it as being tiled by a set of disks of identical radius $L$.  When the disk radius $L$ exceeds the Jeans radius of the disk, the disk is unstable against collapse.  The Jeans radius $L_J$
of a thin disk with surface number density $\Sigma$, comprised of particles of mass $m$ with rms velocity $v_{\rm rms}$, has been computed by
Toomre \cite{Toomre},
\begin{equation}\label{toomre}
L_J= \frac{\pi}{8} \frac{v_{\rm rms}^2}{G m \Sigma}~~~,
\end{equation}
and the corresponding collapse time $t_J$, for $L=L_J$, is
\begin{equation}\label{toomre1}
t_J=\left( \frac{\pi L_J}{8 G m \Sigma} \right)^{1/2} ~~~.
\end{equation}

Combining the previous formulas, and noting that $g/G=M/(R^*)^2$ can be used to eliminate both $g$ and $G$, we get our final formulas
\begin{align}\label{final}
L_J=&\frac{3 \pi}{8} \left(\frac{15}{\pi^2}\right)^{1/4} \frac{M}{(R^*)^2}\frac{c^{1/2}\hbar^{3/4}}{m^{7/4}v_{\rm w}^{1/2}f^{3/4}}~~~,\cr
t_J=&\frac{3^{1/2} \pi}{8} \left(\frac{15}{\pi^2}\right)^{1/8} \frac{M}{(R^*)^2}\frac{c^{1/4}\hbar^{3/8}}{m^{11/8}v_{\rm w}^{3/4}f^{7/8}}~~~.\cr
\end{align}

\subsection{Putting in Numbers}

To reduce our results to numerical form, we define dimensionless numbers $D$, $V$, and $F$ characterizing respectively the wind particle mass $m$, the wind rms velocity $v_{\rm w}$, and the wind flux $f$, by writing
\begin{align}\label{num}
m=&D\, m_{\rm proton}  = D\times 1.672 \times 10^{-27} {\rm kg}~~~,\cr
v_{\rm w} =& V\, c = V \times 3 \times 10^5 {\rm km}/{\rm s} ~~~,\cr
f= &F\, {\rm nanomole}/({\rm cm}^2 {\rm s})= F\times  6 \times 10^{24} /({\rm km}^2 {\rm s})~~~,\cr
\end{align}
together with
\begin{align}\label{consts}
M=& 4.1 \times 10^6 M_\odot = 8.15 \times 10^{36} {\rm kg}~~~,\cr
R^*=&0.42 \times 10^{13} {\rm km}~~~,\cr
c=& 3 \times 10^5 {\rm km}/{\rm s}~~~,\cr
\hbar=& 1.06 \times 10^{-40} {\rm km}^2 {\rm kg}/{\rm s}~~~,\cr
k=&1.38 \times 10^{-29}{\rm km}^2 {\rm kg}/({\rm s}^2 {\rm Kelvin})
\end{align}
Substituting these into the equations for $L_J$, $t_J$, and $T$, we get
\begin{align}\label{formulas}
L_J=&N_1\, D^{-7/4}V^{-1/2}F^{-3/4} {\rm km}~~~,\cr
t_J=&N_2\, D^{-11/8}V^{-3/4} F^{-7/8} {\rm s}~~~,\cr
T=&N_3 D^{1/4}V^{1/2}F^{1/4} {\rm Kelvin}~~~,\cr
\end{align}
with the numerical constants $N_1$, $N_2$, $N_3$ given by
\begin{align}\label{numconsts}
N_1=&1.19 \times 10^{10}~~~,\cr
N_2=&8.47 \times 10^8~~~,\cr
N_3=&7.98 \times 10^3~~~.\cr
\end{align}

From these formulas we can place constraints on the postulated wind parameters $D$, $V$, and $F$.  Requiring that the Jeans length $L_J$
be less than the distance $R^*$ from the black hole to the region containing young stars, we get
\begin{equation}\label{ljcond}
D^{-7/4}V^{-1/2}F^{-3/4} <3.5 \times 10^2~~~,
\end{equation}
and requiring the collapse time to be less than a  few times the estimated age of the young stars, which is  $10^6 {\rm years} \simeq 3.1 \times 10^{13} {\rm s}$, we get
\begin{equation}\label{tjcond}
D^{-11/8}V^{-3/4} F^{-7/8} <3.6 \times 10^4~~~.
\end{equation}

\subsection{Problems arising from neglect of three dimensional aspects}

We now note two problems associated with the model as formulated, both pointed out to us by James Stone.\cite{stone}  Paraphrasing  from his email, the first problem is that the young stars observed near the galactic center are in orbit around the central black hole, and so the disk of gas formed by the wind must acquire enough angular momentum for any newly formed stars to orbit in place, rather than falling back into the hole.
The second problem is that  the volume density of particles in the wind will fall off as the inverse square of the distance from the black hole (if the flux of particles is conserved on spherical shells from the hole to where the disk forms).  This might lead to very high densities near the hole, rasing  the question of whether the mechanism producing the wind can lead to such a high flux at the black hole surface.  We shall see that these two problems with the preliminary formulation of the model are related, with the solution of the first suggesting also a solution to the second.

\subsection{Angular momentum considerations}

For an object in a circular Newtonian orbit around a central mass $M$, with angular velocity $\omega$ at radius $R^*$, the attractive and centrifugal forces must balance.  Using now geometrized units with the Newton constant G (and also the velocity of light $c$) taken as  unity, we have
\begin{equation}\label{bal}
\frac{M}{(R^*)^2}=\omega^2 R^*~~~.
\end{equation}
Rewriting this in terms of the angular momentum per unit mass $j=(R^*)^2\omega$ of the orbiting object, we
arrive at the size of the required angular momentum needed for orbital motion,
\begin{equation}\label{jform}
j=(M R^*)^{1/2}~~~.
\end{equation}

How can this angular momentum be supplied to the disk of gas?   Although the black hole may be rotating, even if the wind particles circulate
with velocity $c$ at the nominal horizon of the hole, they will carry angular momentum per unit mass of order $j \sim 2M$, smaller than the needed angular momentum of Eq. \eqref{jform} by a factor $\sim 2 (M/R^*)^{1/2} <<1$.   So one cannot invoke rotation of the black hole to account for the needed angular momentum.

However, particles far above the disk, even if not having enough angular momentum to be orbital, can still by Eq. \eqref{jform} have significant levels of angular momentum that can be transferred to the disk by collision and absorption into the disk.  Therefore, the preliminary assumption that we made, of neglecting a flux of infalling particles, must be dropped in order to account for the disk obtaining
the needed angular momentum to be in orbit around the black hole.

\subsection{Modification of the model with infalling particles supplying the angular momentum}

If there are infalling particles, Eqs. \eqref{incr}, \eqref{supp}, and \eqref{enbal1} will be modified to take into account
the infalling mass, momentum, and energy, in addition to the infalling angular momentum.  The effects of the mass and energy balance
modifications will be qualitative but not decisive.  However, if enough angular momentum is transferred to the disk for it to become orbital, then it is no longer necessary for the upward wind pressure to support the disk from falling into the hole; at most one might ask that the upward wind pressure balance the downward infalling particle pressure on the disk.  This is a much weaker requirement on the wind flux than that it balance the disk weight, and will permit the wind flux to be smaller than our preliminary estimates, potentially solving the second problem noted above.  Another consideration relevant to the second problem is whether the wind postulated in our model is
a steady-state, $4\pi$ steradian phenomenon, or is a transient eruption of limited angular extent.

\section{Discussion}

These results suggest that our model may be viable with reasonable values for the the black hole wind parameters.   To say more will require detailed modeling of the wind parameters, as well as of the parameters associated with infalling particles.

To conclude, we note that our model may usefully complement the calculation of Bonnell and Rice \cite{bonnell}, who have numerically simulated a giant molecular cloud interacting with the galactic center black hole.  They find that part of the cloud becomes bound to the hole, leading to a
disk that rapidly fragments to form stars.  At the end of their paper they note that although their model is promising, ``What is still unclear,
however, is the origin of the infalling cloud and the probability of the small impact parameter that is required''.  The one dimensional model that we have formulated above supplies possible answers to these questions, since if the cloud forms from a particle wind emanating from the black hole interacting with infalling particles, the impact parameter can be small, and conditions for relevance of the simulation of \cite{bonnell} may be satisfied.

\section{Acknowledgements}

We wish to thank James Stone for reading the preliminary version of this paper and raising the pertinent questions discussed above.

 \begin{figure}[t]
\begin{centering}
\includegraphics[natwidth=\textwidth,natheight=300,scale=1.8]{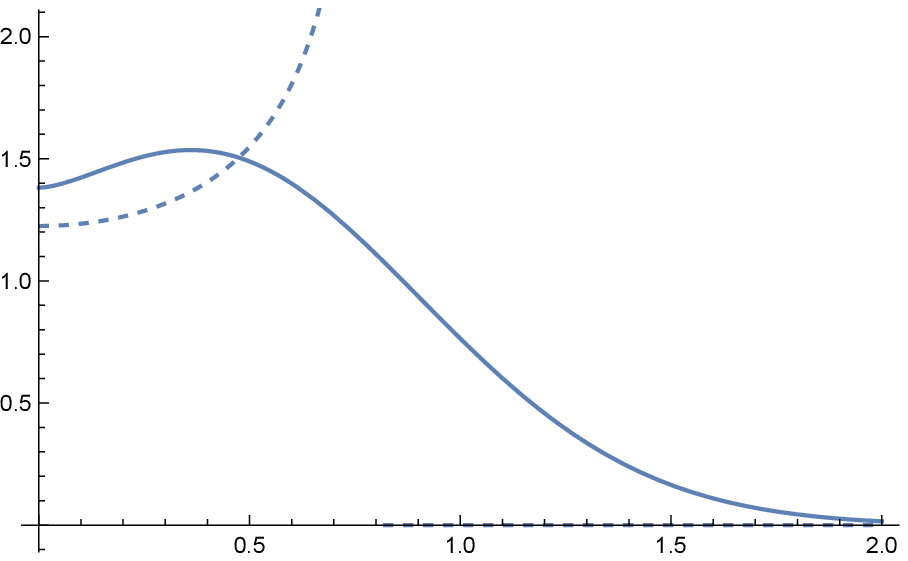}
\caption{Plots of the cusp density of Eq. \eqref{dens} (dotted line) and of the smeared density of Eq. \eqref{smeared2} (solid line) on the vertical axis,
for the illustrative parameter values $f=1$ and $B=1.5$, versus $A$ along the horizontal axis. }
\end{centering}
\end{figure}

\end{document}